# Electricity generation from renewable energy based on abandoned wind fan

Arni Munira Markom[1], Muhammad Hakimi Aiman Hadri[1], Tuah Zayan Muhamad Yazid[1], Zakiah Mohd Yusof[1], Marni Azira Markom[2,3], Ahmad Razif Muhammad[4]

[1]School of Electrical Engineering, College of Engineering, Universiti Teknologi MARA, Johor, Malaysia
[2]Faculty of Electronic Engineering Technology, Universiti Malaysia Perlis, Perlis, Malaysia
[3]Advance Sensor Technology, Centre of Excellence, Universiti Malaysia Perlis, Perlis, Malaysia
[4]Institute of Microengineering and Nanoelectronics (IMEN), Universiti Kebangsaan Malaysia, Selangor, Malaysia

**ABSTRACT**

In the 21st century, our world is facing difficult conditions for serious environmental pollution and the problem of energy shortage. An innovative idea has emerged to recycle wind energy from air conditioning condenser fans in outdoor buildings. Therefore, the main goal of this research is to develop renewable wind energy from the condenser fan of an air conditioner using Arduino as a microcontroller. This research moves towards a portable, low cost, environmentally friendly mini device that harnesses renewable energies with endless resources for future alternative power generation and reduces the burden of consumers' electricity bills.

*Corresponding Author:*

Arni Munira Markom
School of Electrical Engineering, College of Engineering, Universiti Teknologi MARA
Johor, Malaysia
Email: arnimunira@uitm.edu.my

## 1. INTRODUCTION

A renewable energy is energy produced from natural sources to replace the shrinking and dangerous fossil fuels. Ahmad *et al.* stated that over 90% of fossil fuels are greenhouse gases released through carbon dioxide emissions [1]. The increasing concentration of greenhouse gases is what causes the temperature rise in the atmosphere, known as global warming. Then this hot temperature will melt the polar ice, causing sea level rise on land and climate change worldwide [2], [3]. As a result, renewable energy has received tremendous attention in most countries for its environmentally friendly, reasonable cost, abundance of natural sources, efficiency, and practical uses [4]–[6]. Even if the system faces many challenges to be used on a large scale, especially in the third country, and limited technology experts, this renewable energy emerges over the years. Wind source is one of the renewable energies. Wind is an occasional source of energy that, due to its fluctuating nature, hinders the system, but is still a sustainable and natural source [7], [8]. This energy is required by a wind turbine in order to generate kinetic energy from the rotation of the turbine. Then it is converted into electrical energy by a converter, which can be stored and used for many electrical appliances. Compared to other energy sources, a wind turbine has the lowest impact on the environment because it can avoid air or water pollution from emissions from the water-cooling process [9]. As a result, it can generate safe electrical energy without the need for fossil fuels by reducing the severe air pollution and the release of carbon dioxide.

The consumers of air conditioners (air conditioners) in homes or large buildings are increasing due to the hot weather in Malaysia. The free and abandoned wind source from the air conditioner condenser fan outside the home or building is something we can offer to add more benefit to renewable energy use. In addition, large consumers of air conditioning systems suffer from high electricity bills, especially in Malaysia, due to the high-power consumption of the air conditioning system itself. Meanwhile, traditional wind systems suffer from a huge and heavy wind turbine and high acquisition costs for the construction of the wind turbines.

There are many renewable energy projects around the world that use wind as a natural source of electricity, for example in Poland, India, Kenya, Brazil and Netherlands [10]–[14]. China is now the largest producer of wind energy [15]. The main reason for these countries to recycle wind as energy because they have large open land areas, they can by and large implement their gigantic wind turbines on their land. In this article we will only focus on writing a literature review based on Arduino for your wind turbine projects. In the paper, Mahmuddin *et al.* developed a wind generator that uses Arduino as a microcontroller and is able to control the overload voltage [16]. The charge controller with controller relay to increase the heating and power generator safety during the charging process. However, this application of the charge controller is unsuccessful because of the strong stability of the power delivery of the alternating current (AC) ventilator, and therefore the system does not exceed the charge voltage.

A small Arduino MEGA-based wind power generation system was developed by Mubarok *et al.* using a 3-blade wind turbine [17]. A rotating rotor blade speed, wind speed, wind direction and voltage generated by the direct current (DC) generator were measured and analyzed. The mechanical energy from wind turbines was then converted into electrical energy and stored in a battery. An inexpensive and portable electric generator that uses wind and water sources and is used solely as a replacement for a power bank to charge a cell phone, by Ramli *et al.* [18]. A wind and water generator were used to generate electricity to charge the cellphone with an Arduino as the project's microcontroller. A liquid crystal display (LCD) was used to show the voltage generated when the cell phone was being charged by the system.

In the paper, Patil *et al.* [19] designed a hybrid power generation with an Arduino as a microcontroller using wind and water energy for rural agricultural areas, as shown in Figure 1. The wind and waterpower will turn a single turbine and then convert the energy sources into electrical energy that is stored in a battery. A global system for mobile communication (GSM) module is connected to the Arduino to send soil moisture information and enable control of the on- and off-water controller system via smartphone. However, the turbines cause noise pollution and the system are also inefficient and not very stable, without mechanical components being used in real operation. Another hybrid energy project based on Arduino Uno with wind and photovoltaics is being carried out by Restu *et al.* [20] demonstrated. The Arduino is used to synchronize pulse width modulation (PWM) output, define battery capacity, control wind and photovoltaic systems, and display information on the liquid crystal display (LCD). A voltage sensor is used to determine the output voltage, a relay to control the flow of current and a buck converter circuit to reduce the voltage to the desired level.

Therefore, the main goal of this project is to develop a portable and inexpensive renewable energy capable of generating electricity from a wind source from an air conditioning condenser fan outside the user's home or buildings. An Arduino MEGA serves as the brain of the project to control the renewable energy system. This Arduino is a formidable microcontroller that has successfully developed many technology products under development such as robots, intelligent farming systems, and internet of things (IoTs) based projects [21]–[26]. In addition, the second goal is to use this renewable energy of power generation for small household appliances. The electricity is transmitted to the house or buildings to reduce the electricity bill of the users who are appropriate, such as. B. Cell phones, power banks and many small electronic devices.

## 2. METHOD

Illustratively, Figure 1 is a flowchart showing the wind electricity generation system flowchart based on condenser fan air-conditioning wind sources. When the air-conditioning is turned on, the wind from the condenser fan is spinning at outside home and buildings. Then the wind will move the prototype's turbine which is connected to the Arduino system. It is converted from kinetic energy to electrical energy and sent to a DC motor generator to store and charge a battery. Once the battery is at full capacity, a load such as a small electrical appliance can be connected to provide the power. A liquid crystal display (LCD) shows either the battery when the air conditioning is off and the wind turbine when the air conditioning is on. The system will turn off when the air conditioner is turned off.

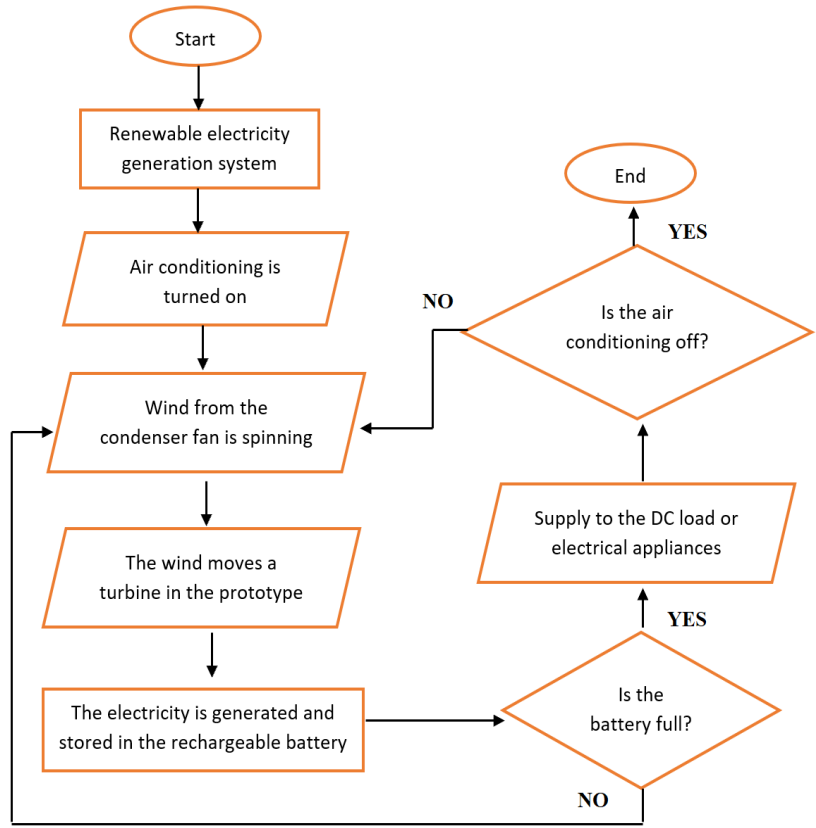

Figure 1. Wind electricity system flowchart

Figure 2 shows the block diagram of the system. The input is the wind from an air conditioning condenser fan that is passed through a DC motor generator. This generator is connected to an Arduino MEGA, which acts as a microcontroller to power the entire system. Then the output consists of light-emitting diodes (LEDs), which are used to define charging (green LED) and to stop charging (red LED), and a USB output as a connection to electrical devices such as mobile phones, power banks and lamps.

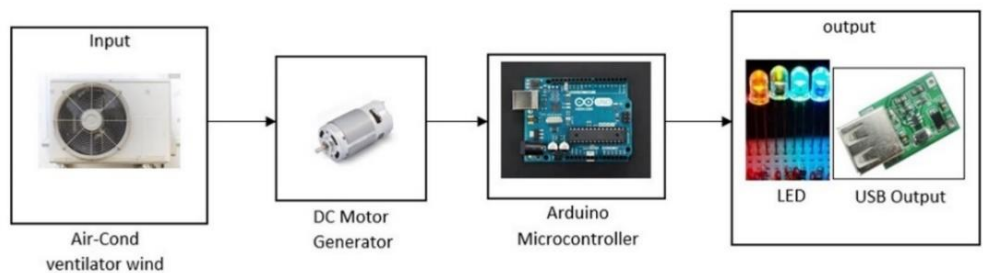

Figure 2. Block diagram of the system

Figure 3 is the software code for the wind turbine. In this code, int input is used to read the Arduino analog pin that is connected to analog pin A1. The float fan for calculating the electricity produced by the wind turbine. An option if and if you can choose a fan with more than 1.5 V or less. The first option with 1.5 V for determining the air conditioning is switched on by a user. When the wind turbine is then started to generate power from the air conditioning fan, the kinetic movement sends a notification to Arduino to display wind turbine on the LCD display. This indicator indicates that the operating system is currently using electrical energy from the wind turbine or the air conditioning condenser fan. The second choice is when the air conditioning is off in the user's home. The Arduino will not detect any power being generated and will notify the LCD display with the word Battery, indicating that the USB charger and LED are currently running on power from the battery.

Figure 4 shows the circuit of the simulation project. For the LCD connection to the Arduino, the LCD pin Vss is connected to ground on the Arduino and the Vdd pin on the LCD which is connected to the 5 V source on the Arduino to provide power to the LCD. Pin RS, RE, D4, D5, D6, and D7 on the LCD connected to the Arduino to control the output that will be displayed on the LCD. In the meantime, connection for wind turbine to Arduino Uno and output LED and USB charger. It can be seen from the diagram that the 5 V battery is directly connected to the normally closed relay that is connected to the output LED and USB charger. When the wind turbine is in operation, the Arduino will detect the current through digital input A1 and automatically activate the relay to switch to the normally open and output LED and the USB charger will be powered by the wind turbine.

```
void loop() {
  int input = analogRead(analogPin);    //read input from wind turbine
  float fan = input * (5.0 / 1024.0);   //input voltage calculation
  lcd.clear();
  if (fan >= 1.5){                      //if wind turbine voltage higher than 1.5v
    digitalWrite(8, HIGH);              //send signal to relay - normally closed
    lcd.setCursor(0, 0);
    lcd.print("Wind Turbine");          //print wind turbine at lcd
    delay(10);
  }
  else{                                 //if no voltage from wind turbine
    digitalWrite(8,LOW);                //not send signal to relay - normally open
    lcd.setCursor(0, 0);
    lcd.print("Battery");               //print battery at lcd
    delay(10);
  }
  Serial.println(fan);                  //print wind turbine voltage value at serial monitor
  delay(500);                           //delay for 0.5 seconds
}
```

Figure 3. Coding for wind turbine

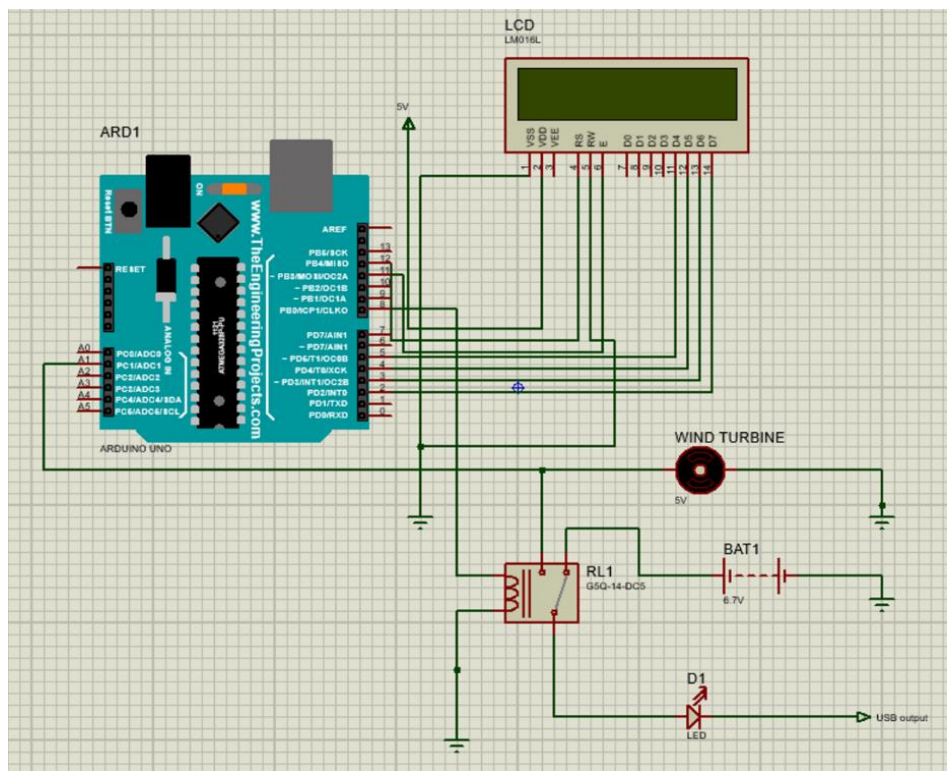

Figure 4. Simulation project circuit

## 3. RESULTS AND DISCUSSION

The idea of this project is to use a wind source, a kinetic energy from the condenser fan of an air conditioner that is normally abandoned outside of homes and buildings. A permanent capacitor working as a generator is connected to a wind turbine and enables the conversion of kinetic or mechanical energy into electrical energy. Then the current is sent directly to a charge controller to manage the voltage and power an Arduino and DC load. The inputs and outputs of the DC load are controlled by the Arduino microcontroller as shown in Figure 5. The charging voltage is then stored in a rechargeable battery.

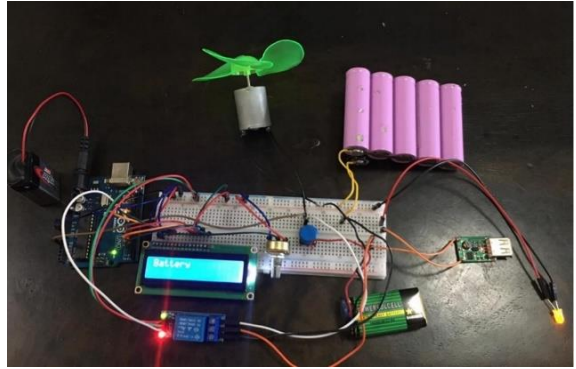

Figure 5. Prototype of the proposed project

Figure 6 is an LCD display to show the notification to the user. Figure 6(a) shows the condition when the air conditioner is switched off by the user, the wind turbine does not work and the LCD shows 'Battery', which is indicated by the output LED and the USB charger that is currently being fed by the battery. Meanwhile, Figure 6(b) shows the condition when the air conditioner is on, the wind turbine is generating power, and the LCD displays 'Wind Turbine', which indicates the output LED and USB charger that are currently powered by the wind turbine. Figure 7 shows the implementation of the prototype on a real air conditioning fan. The prototype was developed to enable mobility without affecting the functionality of the wind turbine. The front of the prototype shows the LCD display, while the back of the prototype has the wind turbine to rotate and generate electricity.

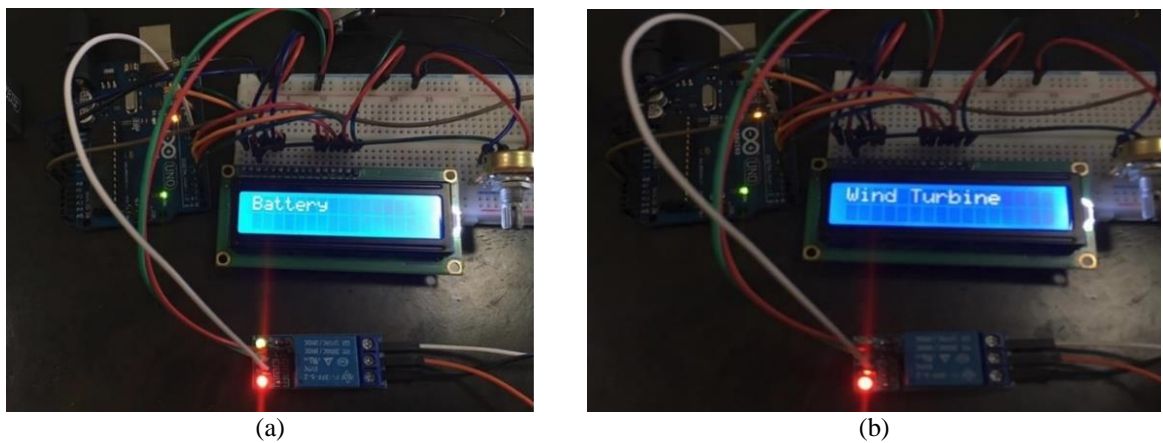

(a)　　　　　　　　　　　　　　　　　　(b)

Figure 6. The LCD display: (a) when wind turbine is not operating and (b) when wind turbine is operating

Table 1 is the results of time versus voltage charging at 10-minute intervals 12 V is generated after 15 minutes, followed by 12.93 V, 13 V, and 13.04 V every 10 minutes. After 45 minutes, the voltage no longer rises because the maximum electrical charge on this prototype is 13.04 V. These 13.04 V and always rechargeable are sufficient for the use of small household appliances such as lamps, fans, kettles, for charging smartphones, and laptops.

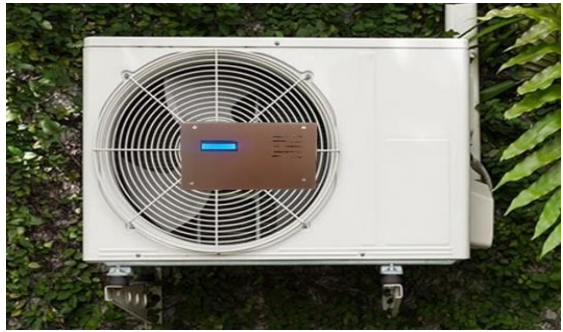

Figure 7. The prototype attached on the air-conditioning ventilator

Table 1. Time taken versus amount of charged voltage
| Time (minutes) | Charged voltage (V) |
| --- | --- |
| 15 | 12 |
| 25 | 12.93 |
| 35 | 13 |
| 45 | 13.04 |

## 4. CONCLUSION

A wind source from an air conditioning fan that was abandoned outside of our home is successfully recycled to generate electricity using an Arduino microcontroller. The maximum voltage generated is 13.04 V at 45 minutes, which is sufficient for small devices at home or in the office. Wind energy is a sustainable, renewable energy and, in contrast to burning fossil fuels, has less of an impact on the environment. This will lead to more research, such as implementing the internet of things (IoTs) and big data on remote controlled and monitored data to replace traditional non-renewable energy for centuries.

## ACKNOWLEDGEMENT
This work is supported by Ministry of Education (MoE) Malaysia under grant 600-IRMI/FRGS-RACER 5/3 (107/2019).

## BIOGRAPHIES OF AUTHORS

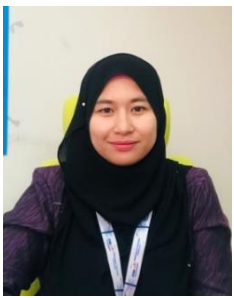

**Arni Munira Markom** 🆔 **g** SC P is Senior Lecturer at the School of Electrical Engineering, College of Engineering, Universiti Teknologi MARA, 81750 Masai, Johor, Malaysia. She received her PhD in Electronics (Photonics Engineering) from Universiti Malaya, Malaysia in 2016. She previously had a Masters in Microelectronics from Universiti Kebangsaan Malaysia and a Bachelor of Electronics (Computer Engineering) from Universiti Teknikal Malaysia Melaka, Malaysia. Her research areas are photonics technology, fiber lasers, fiber sensors and electrical engineering including microcontrollers and IoT devices. She can be contacted at email: arnimunira@uitm.edu.my.

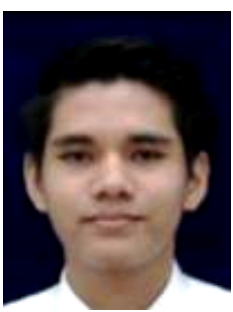

**Muhammad Hakimi Aiman Hadri** 🆔 **g** SC P is graduan from the School of Electrical Engineering, College of Engineering, Universiti Teknologi MARA, 81750 Masai, Johor, Malaysia on last 2021. He received his Diploma in Electrical Engineering major in Electronics. He demonstrates his excellence in science and sports and has a keen interest in Electrical Engineering. Now, he is pursuing his bachelor's degree at Universiti Teknologi MARA in Electrical Engineering. He can be contacted at email: mhkimi1657@gmail.com.

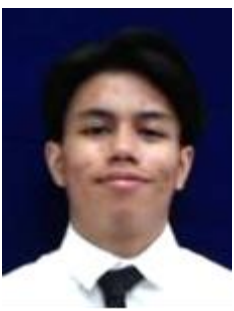

**Tuah Zayan Muhamad Yazid** 🆔 **g** SC P is graduan from the School of Electrical Engineering, College of Engineering, Universiti Teknologi MARA, 81750 Masai, Johor, Malaysia on last 2021. He received his Diploma in Electrical Engineering major in Electronics. Recently, he continued his studies in Bachelor degree in Electrical Engineering at Universiti Teknologi MARA, Shah Alam, Malaysia. He can be contacted at email: tuahzayan@gmail.com.

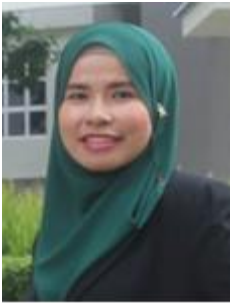
**Zakiah Mohd Yusoff** is a Senior Lecturer at the Faculty of Electrical Engineering, Universiti Teknologi MARA Kampus Pasir Gudang Johor, Malaysia. She received her PhD in control systems from the Faculty of Electrical Engineering, University Teknologi MARA Shah Alam, Malaysia in 2014. Previously, the bachelor's degree of Engineering B.A. Awards was obtained in 2009 from the Universiti Teknologi MARA. She has authored or co-authored more than 25 journals, 35 proceedings, with 8 H-index in Scopus. Her research interests include control system, system identification, modeling, artificial neural network (ANN) and internet of thing (IoTs). She can be contacted at email: zakiah9018@uitm.edu.my.

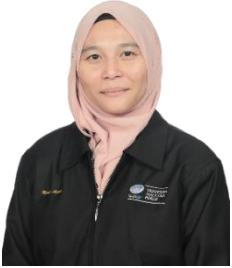
**Marni Azira Markom** received the PhD, MSc., BEng. (Hons.) in Biomedical & Computer Engineering from Universiti Malaysia Perlis. Her research area is Mobile Robot Localisation, Signal Processing & Artificial Intelligence, Recycling System. She has more than 35 journals and 5 H-index continuous publications in Scopus. She is a Senior Lecturer in the Department of Electrical Engineering at Universiti Malaysia Perlis and a member of Advance Sensor Technology, Center of Excellene at the same university. She can be contacted by email: marni@unimap.edu.my.

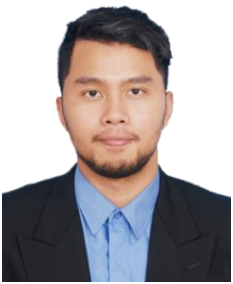
**Ahmad Razif Muhammad** received his B.Eng. (Hons) in Mechanical Engineering from the University of Malaya; M. Phil in Photonics Engineering from the University of Malaya; and PhD in Electrical Engineering (Photonics) in University of Malaya. His doctoral thesis focuses on short pulse fiber laser based on pure metal saturable absorber. Currently a research fellow at Institute of Microengineering and Nanoelectronics (IMEN), Universiti Kebangsaan Malaysia (UKM), Malaysia. His research interests include NIR fiber laser, ultrafast fiber laser, nanofiber for optical sensing. He can be contacted by email: a.razif@ukm.edu.my.